Evolving Paradigms in Task-Based Search and Learning:
A Comparative Analysis of Traditional Search Engine with LLM-Enhanced
Conversational Search System

Zhitong Klara Guan, Yi Wang

# 1 INTRODUCTION

As digital technologies continue to advance, search engines remain pivotal in facilitating access to and interaction with information. The emergence of Large Language Models (LLMs) represents a significant milestone in information retrieval (IR), showcasing unparalleled abilities in understanding, generating, and inferring knowledge from vast textual datasets. Traditionally, keyword-based queries have dominated academic and general search landscapes, albeit with notable limitations. Research by Hassan Awadallah et al. (2014) indicates that nearly half of all web search queries go unanswered, underscoring the challenge in addressing complex and multi-step information needs using conventional search methods. This gap in search efficacy underscores the necessity for more advanced tools capable of navigating the complexities of modern information-seeking tasks.

LLMs, such as ChatGPT and Claude, represent a breakthrough in addressing these challenges. Built upon foundational models like GPT-4 and augmented with specialized skills and plugins, these LLMs offer nuanced understanding and sophisticated generation capabilities that transcend traditional search parameters. They are specifically engineered to optimize search results for intricate queries, providing support for creative and multifaceted tasks previously beyond the reach of standard search technologies. The integration of LLMs into mainstream search platforms signifies a paradigm shift in online information retrieval, with leading search engines like Google and Bing leveraging LLM technologies to enhance user search experiences.

Despite the extensive focus on investigating people's lookup tasks, research on complex search scenarios, such as exploratory search, remains limited. For tasks involving exploratory search and search-based learning, the impact of LLMs on users' search processes and learning outcomes remains largely unexplored.

In light of this, our study delves into individuals' search behaviors and learning outcomes within both regular search systems and LLM-powered search systems. The primary objectives of our investigation are to (a) explore the search process within LLMs and (b) assess the impact of LLMs on learning outcomes in search-based learning tasks. Based on that, we have formulated two research questions to guide our investigation:

**RQ1. How do individuals' search behaviors within Large Language Model (LLM)-powered search systems differ from those within regular search systems?**
Specifically, we aim to explore variations in search strategies, query formulation processes, and information evaluation methods between these two search environments.

**RQ2. What is the impact of utilizing LLM-powered search systems on learning outcomes in search-based learning tasks?**

Our investigation seeks to assess the effects of LLMs on learning outcomes, including factors such as information comprehension, knowledge integration, and the development of critical thinking skills during search-based learning activities.

These research questions will serve as the foundation for our exploration of individuals' search behaviors and learning outcomes within diverse search environments. Through rigorous examination and analysis, we aim to contribute valuable insights into the evolving landscape of information retrieval and its implications for learning processes.

## 2 RELATED WORK

### 2.1 Exploratory Search & Comprehensive Search

Search activities are broadly classified into two main types: Lookup and Exploratory Search (Marchionini, 2006). While lookup search aims for quick and precise results of a "known item'", exploratory search represents a multifaceted, open-ended, and persistent information-seeking context, characterized by strategies that are iterative, opportunistic, and employ various tactics, making it a complex area for both users and the designers of information retrieval systems (White & Roth, 2009). Marchionini (2006) summarized that essential elements of the exploratory search process involve both learning and investigation.

With increasing discourse in academia surrounding "search as learning" in exploratory search, Rieh et al. (2016) introduced the concept of "comprehensive search", which describes "iterative, reflective, and integrative search sessions that facilitate critical and creative learning beyond receptive learning". Comprehensive search encompasses diverse search activities that foster critical thinking and creativity. Unlike exploratory search, which focuses on tasks and outcomes, comprehensive search emphasizes the learning process during search activities, intertwining cognitive learning modes with search behaviors. While both explore similar activities like knowledge acquisition and interpretation, comprehensive search uniquely aims to support not only critical learning but also creative learning, setting it apart in the realm of information retrieval and cognitive development.

### 2.2 LLM & Information Seeking

Initial endeavors in integrating Large Language Models (LLMs) into information retrieval (IR) systems have demonstrated their potential to significantly enhance search experiences. For example, ChatGPT has showcased remarkable capabilities in addressing information-seeking needs through an understanding of user queries. Similarly, the New Bing aims to improve search utility by synthesizing information from various web sources into succinct summaries that directly answer users' queries (ZHU 2024). Additionally, LLAMA-index represents an advancement as a Retrieval Augmented LLM, furthering the effort to leverage LLMs in search solutions.

A key advancement in these systems is their inclusion of reference features, which was a notable limitation in earlier LLM-powered tools like ChatGPT. Recent developments, such as in perplexity.ai3, Bing Copilot and Google Bard, have prioritized the ability to provide references. Highlighted in studies by Nakano et al. (2021) and Menick et al. (2022), these commercial platforms have started responding to user queries with natural language answers accompanied by references to web pages.This enhancement not only improves information credibility but also underscores the evolving role of LLMs as crucial facilitators of information retrieval and knowledge discovery.

## 2.3 Model of Information Search Process (ISP)

Kuhlthau's information search process model (Kuhlthau 2005) stands as one of the most classical models in the field of IR, offering an analytical framework to understand individual's search behaviors. The Information Search Process (ISP) provides a thorough perspective on how users seek information. It outlines six main stages: task initiation, selection, exploration, focus formulation, collection, and presentation. Across these stages, the ISP considers three aspects of user experience: emotions, thoughts, and actions, which remain consistent throughout. This comprehensive framework provides researchers and practitioners with a nuanced understanding of how individuals navigate the information landscape, containing not only the cognitive processes involved but also the affective and behavioral aspects of search behavior.

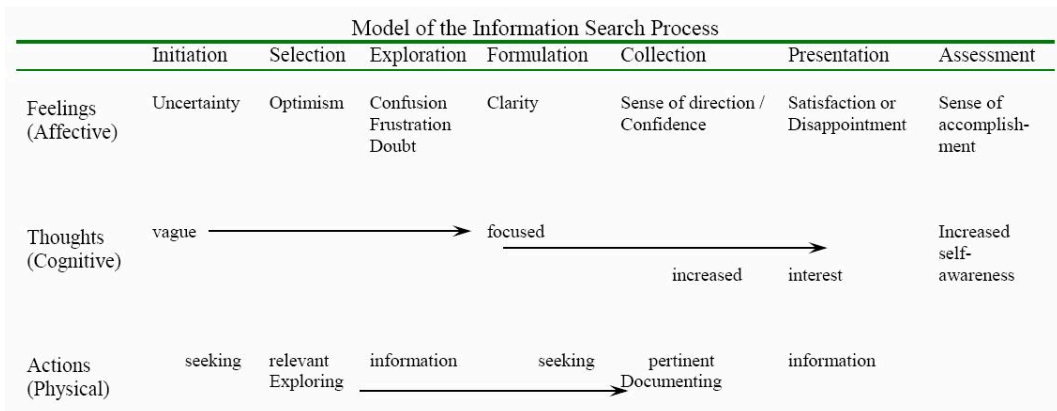

Figure1: Model of Information Search Process (Kualthau, 2005)

The ISP framework evolved over two decades of empirical research, starting with qualitative studies of secondary school students (Kualthau 1997, Kracker 2002, Kuhlthau 2008) and culminating in diverse investigations involving library users and workplace scenarios (Kuhlthau, 2001). Findings from these studies reveal that early stages of information seeking are marked by emotional symptoms like uncertainty, confusion, and frustration, often stemming from vague thoughts about a topic or problem. As thoughts become clearer and more focused, feelings of confidence and certainty tend to increase. Emotional factors, such as uncertainty and confusion, are as influential as cognitive factors, like personal knowledge and information content, in determining relevance judgments. Uncertainty is a central concept in the ISP model, highlighting its significance for intermediaries and system designers, particularly during the exploration stage.

## 2.4 Evaluation of Search Systems
### 2.4.1 Evaluation of Interactive search systems
In the domain of interactive information retrieval (IIR) and exploratory search, four primary metric categories are commonly employed in the user experience measurements: Contextual, Interaction, Performance, and Evaluative Feedback derived from subjective assessments (Kelly, 2009). While contextual and performance measures analyze subjective attributes and system functionality independently, interaction and evaluative feedback measures primarily focus on the interaction between the user and the system. Specifically, interaction metrics are based on users' search behavior and corresponding log data. Some frequently utilized metrics include "number of queries, number of search results viewed, number of documents viewed, number of documents saved, and query length".

### 2.4.2 Evaluation of learning and creativity in search
In exploratory search and research on "learning and search", learning constitutes a pivotal aspect of individuals' search outcomes. As such, researchers advocate for studies that capture learning assessment at the intersection of interactive information retrieval and learning sciences. Studies conducted by Collins-Thompson et al. (2016) and Ghosh et al. (2018) underscore the pivotal role of search interaction in shaping individuals' learning outcomes. These investigations employ diverse methodologies, including the analysis of search logs, learning responses, interviews, and questionnaires, to explore the intricate relationship between search behavior and learning achievements. While Collins-Thompson et al. delve into various tasks and query strategies in web search, Ghosh et al. focus on search behavior and learning outcomes in learning-related tasks. Both studies adopt the Anderson and Krathwohl taxonomy of learning (Anderson et al. 2001) as a framework for evaluating participants' cognitive levels of learning, utilizing its hierarchical structure to subjectively assess the depth and complexity of learning attained through search activities.

Additionally, Urgo (2022) summarizes nine commonly employed methods for measuring learning: self-report, implicit measures, multiple-choice questions, short-answer assessments, free-recall exercises, sentence generation tasks, mind-mapping activities, argumentative essays, and summaries. These varied measurement and assessment methods are tailored to different task implications, enabling researchers to gain nuanced insights into individuals' learning outcomes across diverse contexts.

## 3 METHOD
### 3.1 Participants

To examine search and learning processes for comprehensive search tasks, we conducted an exploratory study involving undergraduate and graduate students. We recruited a total of five participants, comprising two undergraduates and three graduate students from universities across the United States. Participants were selected through a screening process to ensure prior experience with LLMs and frequent use of search engines for learning purposes.

## 3.2 Study Design

We conducted an exploratory study to compare user behaviors on traditional search engines and LLM-enhanced search systems during task-based information seeking. Utilizing a within-subjects experimental design, participants performed search tasks under two conditions: 1) using the LLM-enhanced search system, Bing Copilot, which integrates chat and search functionalities and represents state-of-the-art LLM-enhanced systems; and 2) using a traditional search engine, Bing, for fair comparability with Bing Copilot.

To prevent systematic bias, each participant was assigned with a subject number randomly. The assignment of system order is based on their subject numbers: odd-numbered subjects began with Copilot followed by Bing, while even-numbered subjects started with Bing followed by Copilot.

**Task A:** This semester you are taking a public health class. Yesterday a classmate's post on the course page discussed urban green spaces' effects on mental health. This topic piques your interest and you decide to delve deeper.

**Task B**: Your class is exploring technological opportunities to transform future education. Yesterday a classmate's post on the course page discussed applying robotics in K-12 education. This topic piques your interest and you decide to delve deeper.

Figure2: Task A & Task B Description. Instructions Cropped

To mitigate the learning effects typical in within-subject designs, we developed two tasks of similar domain difficulty, referred to as Task A and Task B (see Figure2). Tasks should be comprehensive search tasks consisting of search, critical and creative learning They should require participants to explore open-ended, niche topics likely unfamiliar to them, necessitating the synthesis of creative, original ideas. Participants received task instructions in a workable Word document, where they could note their findings and craft responses. The instructions were as follows: For the critical learning phase, "Search for information on the topic using the designated system and document your findings here." For the creative learning phase, "Based on your research, write a 4-6 sentence original response to be shared in a Canvas discussion, aiming to contribute unique insights."

### 3.3 Data Collection

Data collection began with the participants filling out a demographic questionnaire followed by an introduction to Task A via the prepared document. A pre-task questionnaire assessed their prior knowledge, perceived task difficulty, and interest in the subject. Participants were then randomly assigned one of the two search conditions and asked to perform three initial searches (if Bing) or AI interactions (if Copilot) to familiarize themselves with the system. If the search system was Bing, participants were instructed to turn off "Copilot response on result page" and "Scroll to open Copilot". The participants were then instructed to start working on the search task. Following the task session, a post-task questionnaire evaluated their perceived learning outcomes (Collins-Thompson et al. 2016) and creativity support (Carroll and Latulipe 2009). This was complemented by a semi-structured interview to collect qualitative feedback on user experiences and task effectiveness. The entire process, including the pre-task questionnaire, search session, post-task questionnaire, and interview, was repeated for Task B using the alternate search system. Figure 3 shows the step-by-step procedures we followed.

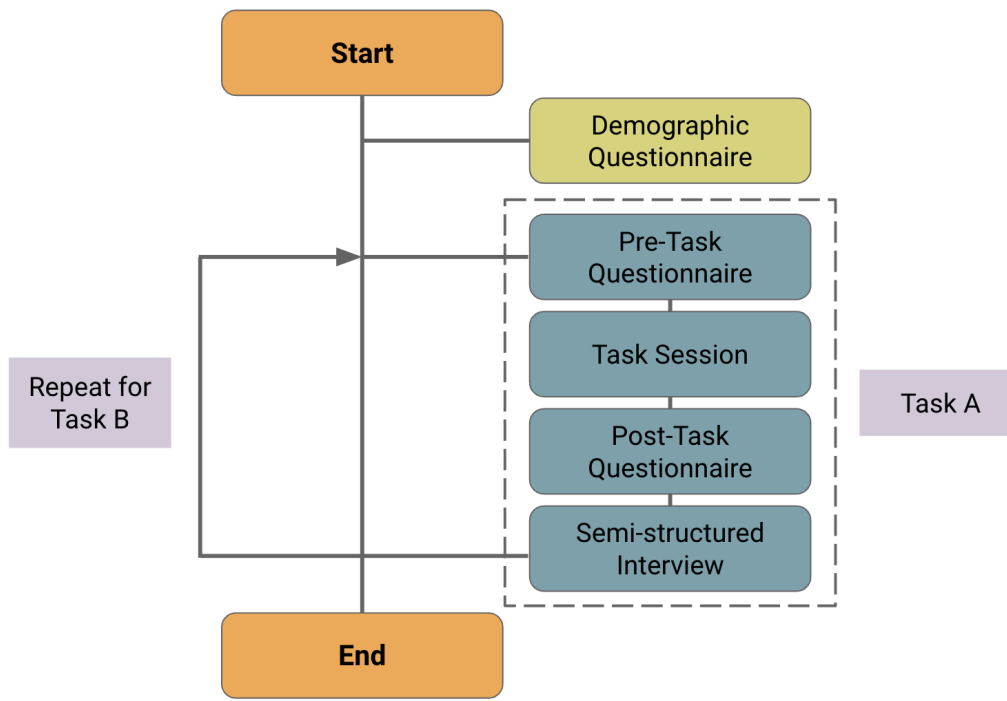

Figure3: Data Collection Procedure

The study was conducted both remotely, using Zoom video conferencing software, and in-person. For all participants, the entire study session lasted approximately 30 minutes. Task sessions were screen-recorded to analyze time spent, query formulation, interactions with search results, and navigation patterns. Interviews were transcribed for further qualitative analysis.

### 3.4 Data analysis

The data analysis for our study comprised three distinct parts:

**Metrics-Based Search Behavior Analysis**: We conducted a comprehensive analysis of search behaviors using metrics derived inductively from the screen recordings. This analysis included data on query or prompt interactions (such as the number, length, and duration of queries, and clicks on suggestions), browsing behaviors (such as maximum scroll depth, cumulative clicks, and duration of page visits), and learning behaviors (such as instances of copying and pasting, and text organization).

**Questionnaire Analysis**: To evaluate and compare user experiences across the two search systems, we analyzed the data collected from pre-task and post-task questionnaires. Participant ratings were assessed using the Mann-Whitney U test, a non-parametric method appropriate for the ordinal data generated by the 11-point Likert scale used in our questionnaires (ranging from 0 for least to 10 for most). This method was selected to assess the equality of mean ratings between the paired groups, with a significance level set at $\alpha = 0.05$.

**Content Analysis of Learning Notes, Original Responses and Interviews**: We applied content analysis to the documents and interviews to draw empirical connections between user responses and our thematic research findings. Initial steps involved identifying responses directly related to our research questions, which were then categorized. These categorized texts were subsequently coded thematically, aligning with the two primary levels of focus in our research questions: the search process and the learning process.

### 4 Results

### 4.1 Search with Copilot is More Efficient

Search behavior, questionnaire and interview analysis indicates that Bing Copilot significantly enhances search efficiency compared to Bing.

Such increase in efficiency is particularly evident in the reduced need for search task planning that is typically required in comprehensive search tasks (Chavula, Choi, and Rieh 2024). Participants reported a marked ease in initiating searches with Copilot, attributing this to the system's ability to understand and process comprehensive queries. For instance, participants P03 and P04 directly copied the entire task prompt into Copilot to initiate their searches. Similarly, participants P01 and P02 utilized AI-generated prompt suggestions effectively, which facilitated their search process. The qualitative feedback from participants underscores the benefits of Copilot in initiating search in Bing, especially when they are unfamiliar with a subject. Participant P05 emphasized this point, stating, "So I think it's useful when you need idea generation in this case, like you maybe completely don't know about a subject and you search a new subject and this, the AI shows you: oh, here are some points about the subject that you might want to look into later… At least AI gives you a starting point."

Further enhancing the efficiency of Copilot was the significant reduction in the time and effort participants needed to spend browsing for information. The study observed a decrease in both the proportion of time and the number of documents participants reviewed. This reduction in browsing effort is exemplified by the structured responses from Copilot, which streamlined the information assimilation process. Participant P01 remarked on the structured nature of Copilot's responses: "The Copilot response is very structured. I spent much less effort in understanding what's out there. But in the search system (Bing) I need to open each link, select and synthesize useful points…"

Moreover, participants noted a decrease in cognitive load and fatigue when using Copilot compared to traditional methods. Participant P02 highlighted this difference by commenting on the physical and mental effort required: "For Bing search, I felt tired after even looking for an overview or one aspect…" This sentiment was reinforced by observations from participants P02, P03, and P05, who noted that the information density in Copilot's answers was higher, further contributing to a more efficient search process.

## 4.2 Exploration, Formulation, Collection

Upon analyzing the data related to search behaviors, our findings affirm that Kuhlthau's Information Search Process (ISP) remains a relevant model for understanding information seeking in new technological environments. We annotated and categorized all search sessions—both those using Bing and those using Bing Copilot—into the six stages defined by the ISP model. Significant differences in information seeking behaviors were evident between the two search systems during three specific stages in the middle of ISP: exploration, formulation, and collection.

**Exploration Stage**, characterized by an initial search to gain broad knowledge about a topic, showed notable differences between the two systems. During sessions using Bing, participants generally composed shorter queries and engaged in more extensive browsing, either by scanning multiple documents or deeply reading single results. The queries in these sessions were less varied, which potentially restricted the breadth of exploration. In contrast, sessions with Bing Copilot facilitated easier and quicker exploration, evidenced by the diversity of prompts entered into the system. These ranged from general questions posed to the AI to advancing prompts suggested by Copilot itself, facilitating exploration by covering a wider array of topics more efficiently.

**Formulation Stage** is where searchers develop a focused perspective on their topic (Kuhlthau, 2018). In Bing searches, we observed that as participants' queries became longer and more specific with keywords such as an institution name. Their browsing behavior shifted towards more direct locating strategies, such as using text match within search results to find specific information. With Bing Copilot, formulation often began when participants expressed the need for more focused answers. Notably, some participants would interrupt the AI's response generation if they perceived the answers as insufficiently specific, prompting Copilot to refine its outputs more quickly.

**Collection Stage** is gathering information pertinent to the focused perspective (Kuhlthau, 2018). Both Bing and Bing Copilot facilitated extensive use of copy and pasting strategies, but interactions with the gathered information diverged. With Bing, participants actively grouped and labeled information, a process conducive to deeper idea synthesis. This method was more labor-intensive and time-consuming. With Bing Copilot, however, the selection process typically involved highlighting relevant information and excluding less pertinent ideas, which, while quicker, may not encourage as thorough an engagement with the material.

The differences observed across the three stages of ISP suggest that while foundational information seeking behaviors persist, the tools and methods used can significantly influence in which part of the process people are more engaged.

### 4.2.1 Trade-offs in Engagement Across Stages:

In sessions using Bing, participants often faced a dichotomy: they either exhibited less variation in their queries, reflecting a lack of diversity in perspectives, or they utilized much shorter queries, indicating a narrower focus. This pattern suggests a trade-off between exploration and formulation stages, where participants felt compelled to prioritize one due to limited attention resources. Interviews and notes from these sessions confirmed that while exploration and formulation with Bing required more time, the depth and breadth of learning were often compromised.

For example, in sessions where participants (Bing: P01, P02, P03) engaged extensively in initial browsing, they failed to develop a focused interest, as reflected in their notes and responses that displayed a range of perspectives (e.g., pros and cons; teacher vs. student viewpoints). Conversely, in sessions where focus was quickly narrowed (Bing: P04, P05), the diversity of perspectives in notes and responses was less pronounced. However, these participants tended to provide more detailed examples and studies as supporting evidence, indicating a deeper engagement in the collection stage.

The experimental setup of the study may lead participants to predetermine the total amount of attention and effort allocated for each task, which may not fully replicate real-world search behaviors. Nevertheless, this constraint reflects Kuhlthau's observation that searchers often skip the critical stages of exploration and formulation, moving directly from selection to collection (Kuhlthau 2018).

Overall, the questionnaire and interview data indicated higher search support from Copilot. This preference is likely due to Copilot's support in the more challenging stages of exploration and formulation, as identified by Kuhlthau (Kuhlthau 2004). By aiding in these initial stages, Copilot potentially reduces the cognitive load on users, enabling them to allocate their attention more effectively across the search process.

### 4.3 Affective Journeys of ISP

Upon analyzing the search interaction and post-task interviews, participants experienced negative emotions with both the traditional search system (Bing) and the LLM-enhanced system (Bing Copilot). However, the stage of ISP at which these negative emotions occurred differed, revealing two distinct affective journeys associated with each system.

With Bing, participants were in the dip of confidence during the exploration stage. This was primarily due to the extensive synthesis required to integrate information from multiple sources. Participants often engaged in behaviors indicative of frustration or dissatisfaction, such as quickly closing scholarly papers that might have seemed too dense or irrelevant at first glance, searching within documents for specific information, and frequently relying on Wikipedia as a straightforward, familiar source. These actions suggest a struggle to efficiently digest and synthesize the vast amount of information encountered, leading to feelings of being overwhelmed and reducing satisfaction early on in the process.

With Bing Copilot, the formulation and collection stage was where negative emotions predominantly surfaced. Participants expressed frustration with the AI-generated responses, especially in cases lacking specificity and detail, or when their questions were not adequately answered. This dissatisfaction sometimes led to interrupting the generation process, constantly re-prompting the system, or ultimately turning to traditional search methods as a fallback. Such behaviors indicate a critical expectation gap regarding the quality of information provided by Copilot, highlighting areas where LLM-enhanced systems may need improvement to better support the later stages of the search process.

However, the questionnaire results showed that Copilot received a significantly higher emotional rating in terms of satisfaction, interest, and enjoyment. This suggests that particularly by easing the cognitive load during the initial stages of search, users perceived more position feelings during ISP. This finding shed light on designing search system support for learning and creativity where positive emotions directly led to better outcomes (Langley 2018; Tyng et al. 2017).

## 4.4 Better Perceived Learning Outcomes with Copilot

Participants consistently reported higher scores for perceived learning outcomes when utilizing Bing Copilot compared to the traditional search system. This enhancement is largely attributed to a reduction in uncertainty throughout the exploration process, enabling a more confident and efficient search. Additionally, the ease with which participants could understand and refine their search topics further contributed to their learning.

For instance, Participant P04 observed the distinct advantage of Copilot in making complex information more accessible: "Even if there are creative ideas on the websites, they are not easily accessible, and sometimes I have to read through all of the information … But Copilot makes it easier to understand ideas in an easier way … I experienced this thing with GPT when I read through the really technical kind of thesis like if there was without GPT I wouldn't have been able to understand the thesis." Similarly, Participant P02 highlighted the benefits of Copilot's efficiency, stating, "It's easy to quickly extract information, makes me willing to look for more. Copilot helps me to learn more and generate new ideas by making a high level summary very quickly accessible, save my energy and mental capacity."

## 4.5 The Role of Construction in Learning Outcomes

Despite the technological facilitation provided by Copilot, the data indicates that learning outcomes are significantly influenced by the extent of construction involved in the search process. Participants' deeper learning was associated with more intensive construction activities during their searches, irrespective of the search system used. This aligns with the Construction Theory of Information Seeking, which posits that learning involves an active construction of understanding and meaning from information over time (Kelly 1963).

Reflective comments from participants reinforce this theory. For example, Participant P01 noted the benefits of a more manual search process with Bing: "I felt like I learned much more with Bing because I spent more time looking for information and reading. I memorized some information with Bing." Participant P03 also supported this perspective by emphasizing the cognitive engagement required for deeper learning: "Search has helped me to learn more, it makes me feel like I'm learning more because I have to think for myself, which leads to a deeper understanding. It also left a stronger impression on me with more memorable stuff than the first task with Copilot." In contrast, Participant P04 perceived a decline in learning effectiveness with Copilot, using it more as a writing aid than a tool for in-depth exploration and learning. Meanwhile, Participant P05 found inspiration in AI's ability to provoke curiosity and foster active inquiry.

**5 SOLUTIONS**

Drawing upon our experimental findings, we propose a new information process model and set forth tailored recommendations aimed at enhancing the learning and creativity support of LLM-powered search systems. These initiatives are designed to optimize learning outcomes and stimulate user creativity, paving the way for a more dynamic and responsive search experience.

**5.1 A new information process model in LLMs**
RQ1 investigated whether there are differences in user's information seeking behaviors between regular search systems (Bing) and LLM-powered search systems (Bing Copilot) in exploratory search tasks

Our within-subject experiment suggests that there might be a theoretical shift in people's information seeking models within Kulthul's conceptualized framework. Our hypotheses centered around three primary assumptions:

1. Users may display a tendency to skip the "planning" stage in the information search process, possibly influenced by the advanced capabilities of LLMs to complement people's lack of prior knowledge in that "planning" stage.
2. LLMs might act as facilitators in the exploration and formulation stages of individuals' information-seeking processes, attributable to their highly summarization abilities.
3. Due to a potential tradeoff between exploration and formulation stages and the subsequent collection stage, given users' limited cognitive loads, we hypothesized that individuals, aided by LLMs, might allocate more time to focused collection.

It's important to note that our study, serving as a pilot investigation, was constrained by a limited sample size of five participants. As such, future research endeavors will necessitate broader participant inclusion and more rigorous experimental settings to substantiate and validate these initial observations.

**5.2 Suggestions on how LLM-powered search systems might better supports people's learning and creativity**
RQ2 investigated how LLM-powered search systems might affect people' s learning and creativity

In addition to proposing a new information process model tailored for LLM-powered search systems, our study also puts forth recommendations on how LLM-powered search systems can better support individuals' learning and foster creativity. Our findings underscore the importance of addressing users' need for greater specificity in the information retrieved from LLMs. Despite initial interest sparked by higher-level summaries generated by LLMs, users expressed a desire for more detailed and niche information. Specifically, they seek real-life applications and examples sourced from interview videos, newspapers, and social media platforms to personalize their learning and construct a deeper understanding.

In essence, users are not merely seeking information but engaging in sense-making activities to truly learn and integrate knowledge. However, current LLM systems fall short in meeting these needs. Therefore, we advocate for the development of LLM search system designs that prioritize and accommodate these information needs, thereby enhancing users' learning experiences and facilitating knowledge construction.

To improve the search system's support for individuals' learning and creativity, we propose two solutions:

1. Implement a hybrid search feature that combines the comprehensive indexing capabilities of traditional search engines (like Bing) with the intuitive, context-aware insights of LLMs. This feature would enable users to delve deeper into specific topics by providing options for refining search results based on niche interests or subtopics.
2. Facilitate knowledge integration by providing users with tools to organize, annotate, and synthesize information gathered from LLM-powered search results.

In summary, this pilot study offers valuable insights into the evolving landscape of information seeking within LLM-powered search systems compared to traditional search systems. While our findings suggest potential shifts in information seeking behaviors and offer recommendations for enhancing LLM-powered systems to support learning and creativity, it's important to note the preliminary nature of our investigation. With a limited number of data points, our proposed information process model and recommendations serve as initial hypotheses to be further validated and refined through larger-scale studies. Nonetheless, this study lays a foundation for future research aimed at understanding the intricate dynamics of the LLM-mediated information seeking process and advancing the development of more effective and user-centric search systems.